\documentstyle[graphicx,prl,aps]{revtex}
\begin{document}
%\documentstyle{article} \begin{document} %\begin{center} %{\Large {\bf
%\title{ Kondo tunneling through real and artificial molecules } %}}
%\end{center} \author{Konstantin Kikoin and Yshai Avishai}
%\author{Konstantin Kikoin and Yshai Avishai}
%\address{Department of Physics, Ben-Gurion University, Beer-Sheva 84 105,
%Israel} \date{\today} \maketitle
\title{Kondo tunneling through a biased double quantum dot}
\author{Yshai Avishai and Konstantin Kikoin}
\address{Ilse Katz Center for Nanotechnology,
Ben-Gurion University of the Negev, 84105 Beer-Sheva, Israel}
\maketitle
\begin {abstract} 
Electron tunneling through a system 
formed by two coupled quantum dots in a parallel geometry
is considered within a generalized Anderson model. 
The dots 
are assumed to have nearly equal radii 
but different (and tunable) gate voltages.
In the absence of tunneling to and from the leads, 
the isolated two-dot system (with two electrons 
in it) resembles an 
hydrogen molecule within the Heitler London approximation. In 
particular, it has a singlet ground state and low lying 
excited triplet state. When tunneling is switched on,
and when the gate voltages are properly tuned the 
ground state becomes a triplet and singlet and triplet 
states are intermixed. In the region where charge fluctuations 
are suppressed, the pertinent antiferromagnectic exchange interaction 
has the form $(J^{T} {\bf S} +J^{ST} {\bf P}) \cdot {\bf s}$. 
It is written in terms of the electron spin ${\bf s}$, the 
double dot spin $1$ operator ${\bf S}$ and an additional 
vector operator ${\bf P}$. The operators  ${\bf S}$ and  ${\bf P}$ 
generate the algebra $o_{4}$ of a spin rotator. The related 
Kondo effect is similar to that of a vertical quantum dot, 
discussed and analyzed recently.
\end {abstract} 
\section{Introduction}
Experimental discovery of resonance Kondo tunneling through 
quantum dots under strong Coulomb blockade \cite{Gogo99} is an
impressive recent result in the physics of nanostructures. 
The observation of Kondo-like zero bias anomaly in the tunneling current 
through planar GaAs/GaAlAs quantum dot (QD) with odd electron occupation 
and a net spin $S=1/2$
confirmed earlier theoretical predictions \cite{Glazr88b}. A natural question
then arose whether the Kondo tunneling is possible when the number of electrons
in the dot is even, so that the nominal spin of an isolated QD is zero. Some
experimental data were consistent with the occurrence 
of Kondo resonance also in that case \cite{Rok00a}. Of course, the
simple reason for Kondo scattering in this case is the triplet ground state
which can be realized provided the exchange interaction compensates 
the energy $\delta_e$ of excitation of one electron from the last doubly 
occupied state. Another compensation mechanism is the Zeeman splitting
of excited triplet state in external in-plane magnetic field $B$ 
\cite{APK,Pust00}. 
In this case Kondo resonance in electron tunneling should be observed at a 
specific value of Zeeman energy $E_Z=g\mu_b B=\delta_e$. Unlike the 
conventional Kondo mechanism, this effect arises under a condition of broken
rotational invariance in spin space, and the tunneling induced 
singlet-triplet mixing is shown to be the source of Kondo-like resonance. 
The experimental observation of Kondo tunneling in even Coulomb
blockade windows at $B=1.36$ T in QD formed in nanotube 
\cite{Cobd00} confirmed these theoretical predictions.

Another interesting possibility of realization of Kondo tunneling in QD
with even electron occupation arises when the low-energy part of its 
spectrum consists of spin singlet, spin triplet and singlet charge transfer 
exciton \cite{KA01}. It was shown in the above paper that such spectrum arises
in double quantum dots (DQD) provided this kind of nanostructure consists 
of two wells of different depth and the tunneling is allowed only through the 
"shallow" well with larger radius and smaller Coulomb charging energy $Q$. 
In this case, the singlet-triplet crossover is an {\it intrinsic} property of 
nanoobject in a contact with metallic leads, 
which arises even at zero magnetic field. This artificial molecule
has a natural prototype: the low-energy part of the electron spectrum of 
complex molecules in which a rare-earth ion is secluded in a carbon cage 
has the above mentioned structure provided the covalent chemical bonds exist
between the $f$-electron of a rare-earth ion and $p$-electrons of carbon cage.
Cerocene Ce(C$_8$H$_8$)$_2$ is a known example of such double-shell
molecule \cite{Liu98}. Being adsorbed on a metallic substrate this molecule
is expected to demonstrate the Kondo-type behavior. 
\section{Effective spin hamiltonian for biased DQD}
In the present research we study the tunneling through DQD 
formed by two dots of nearly equal radii in a parallel geometry (Fig.1)
with two separate gates generating voltages $V_g^{l,r}$. 
Such setups were fabricated several years ago \cite{Molen95}. 
\begin{figure}[htb]
\centering
\includegraphics[
height=0.4\textheight,angle=90,
keepaspectratio]{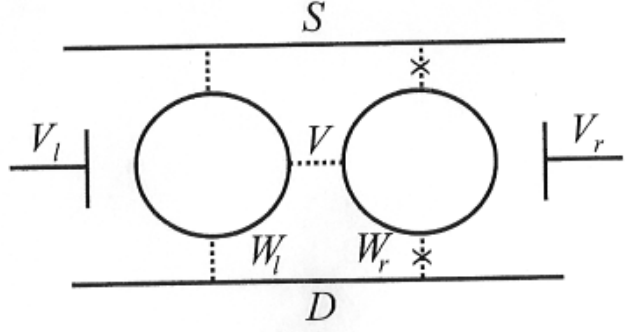}
\caption{Double quantum dot in a parallel geometry.}
\end{figure}
If only one of the 
tunnel channels (say, left) between the dot and the lead is open 
(the inter-dot tunneling $V=0$, the tunnel coupling between the right dot 
and the leads $W_r=0$),
we are left with the "electrometer" configuration when the tunneling through 
the left well is controlled by the charge state in the right well. 
The Coulomb blockade windows between the
resonances in the Coulomb energy of the dot 
${\cal E}_{\nu_r,\nu_l}(V_g^r,V_g^l)$
form a honeycomb pattern where the vertices connect the windows with 
charge configurations $(\nu_r,\nu_l),(\nu_r,\nu_l-1),(\nu_r+1,\nu_l-1)$.
The lines  
${\cal E}_{\nu_l,\nu_r}\approx {\cal E}_{\nu_l+1\nu_r,}$ are the regions where
the Coulomb resonance induced by $V_g^r$ allows tunneling 
through the left dot \cite{Molen95}.

We consider the tunneling through DQD where the source -- drain current  
is allowed only through the left dot, but the tunnel coupling with the right
dot controls the spin degrees of freedom of the DQD. 
The case of DQD occupied by two electrons in its ground state is studied  
assuming that both left and right dots are neutral at $\nu_r=\nu_l = 1.$ 
Then the isolated DQD under strong Coulomb blockade conditions 
reminds, in some respect, the hydrogen molecule in the Heitler-London
approximation. This system is described by a generalized Anderson tunneling 
Hamiltonian 
\begin{eqnarray}
H & = &\sum_{i=l,r}\sum_{\sigma }\epsilon_{i}n_{i\sigma }+
V\sum_{i\neq j}d_{i\sigma}^{\dagger }d_{j\sigma }+
\frac{1}{2}\sum_i Q_i n_i(n_i-1)\nonumber\\
& + & \sum_{k\sigma\alpha}\varepsilon_{k\sigma\alpha}
c^\dagger_{k\sigma\alpha}c_{k\sigma\alpha} +
\sum_{k\sigma}
\left(W_{l}c^\dagger_{k\sigma} d_{l\sigma} + H.c. \right)
\label{1.1}
\end{eqnarray}
Here the quantized energy levels of electrons in the dots 
$\epsilon_{i} =\varepsilon_i + V_g^i$ are biased by 
the gate voltages, the Coulomb blockade energies are assumed to be
the same for both dots, $Q_l=Q_r=Q$,~$\varepsilon_{k\sigma\alpha}$ are
the band energies of the electrons in the leads, 
$\alpha=s,d$ stands for electrons from the source and drain electrodes, 
$c_{k\sigma}=2^{-1/2}(c_{k\sigma,s}+c_{k\sigma,d})$,~ 
$W_{ki}=W_{k\alpha,i}/(W_{ks,i}^2+W_{kd,i}^2)^{1/2}.$ 

We are interested in the case when the artificial Heitler-London molecule is
strongly biased by the gate voltages so that 
${\cal E}_{2,2}-{\cal E}_{1,2}\gg {\cal E}_{1,2}-{\cal E}_{1,1}$, but still
$\beta\equiv V/({\cal E}_{1,2}-{\cal E}_{1,1})\ll 1$ (Fig.2).  
\begin{figure}[htb]
\centering
\includegraphics[
height=0.4\textheight,
keepaspectratio]{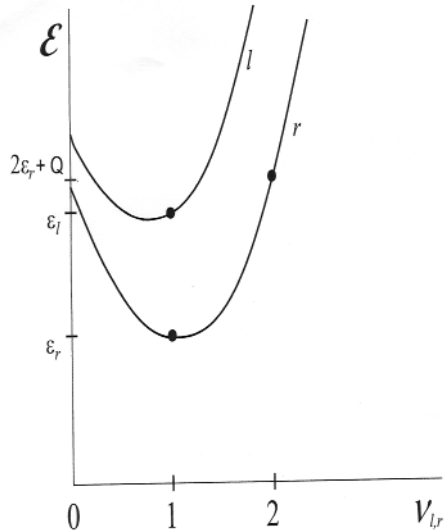}
\caption{Electrostatic energy of the right ($r$) and left ($l$) dot as a 
function of electron occupation numbers $\nu_{r}$ and $\nu_{l}$.}
\end{figure}
In this case 
the dot Hamiltonian (first three terms in eq. (\ref{1.1})) may be easily
diagonalized, and the lowest three levels are 
\begin{equation}
E_S   =   \epsilon_l+\epsilon_r-2\beta V,~~
E_T = \epsilon_l+\epsilon_r,~~
E_R  =  2\varepsilon_r + Q + 2\beta V
\label{2.3} \\
\end{equation}
The small parameter is $\beta=V/(Q-\epsilon_l +\epsilon_r)$ 
expressed in terms of the coupling parameters entering
the Hamiltonian (\ref{1.1}) (see Fig.2). The gate voltage $V_g^l-V_g^r$ 
is applied in such
a way that the spin and charge degrees of freedom are nearly separate in 
isolated DQD, i.e. the energy of charge transfer exciton 
$E_R - E_S = Q-\epsilon_l+\epsilon_r +4\beta V$ substantially exceeds the
energy of spin exciton $E_T -E_S =2\beta V$. 

If the tunnel coupling between the DQD and the metallic leads is also small, 
$W_l/(E_R - E_S) \ll 1$, only the singlet-triplet excitations are 
involved in forming the tunnel transparency of the left dot. In spite of the
occurrence of
singlet ground state, the Kondo-like processes involving the triplet 
exciton are possible, and one can expect the corresponding anomalies 
provided the characteristic Kondo temperature $T_K$ is comparable with
the energy $2\beta V$ of singlet-triplet excitation in DQD \cite{KA01}. 
To study the low-energy anomalies we use the Haldane-Anderson 
renormalization group (RG) approach \cite{Hew93}. According to this method the
Hamiltonian (\ref{1.1}) diagonalized in accordance with (\ref{2.3}) 
is rescaled by absorbing the high-energy excitations of order
$D_0$ (where $D_0$ is the width of conduction band) in the renormalized
parameters of this Hamiltonian at reduced $D=D_0-\delta D$. 
It is known that one can neglect  
renormalization of the tunneling constant $W_l$ in comparison with 
rescaling of the energy levels $E_S$ and $E_T$. The scaling equations
are 
\begin{equation}
dE_\Lambda/d\ln D=\Gamma_\Lambda/\pi,
\label{3.6}
\end{equation}
where $\Lambda=S,T$. The tunnel coupling constants 
$\Gamma_\Lambda=\pi\rho_0|\langle 
0|c_{k\sigma}d_{k\sigma'}|W_l|\Lambda\rangle|^2$ 
are {\it different} for singlet and triplet states because the singlet
exciton ($\rho_0$ is the density of states in the leads). 
In other words, the doubly occupied right dot state 
$|ex_r\rangle = d_{r\uparrow }^{\dagger }d_{r_\downarrow }^{\dagger }|0\rangle$
is admixed to the bare singlet 
$|s\rangle =\frac{1}{\sqrt{2}}\sum_{\sigma }\sigma 
d_{l\sigma }^{\dagger}d_{r\bar\sigma }^{\dagger }|0\rangle$ by inter-dot
tunneling, whereas the triplet states 
$|t_{0}\rangle =\frac{1}{\sqrt{2}}
\sum_{\sigma }d_{l\sigma }^{\dagger}d_{r\bar\sigma }^{\dagger }|0\rangle~,
|t_{\sigma }\rangle =d_{l\sigma}^{\dagger }
d_{r\sigma }^{\dagger }|0\rangle$
are not affected by this tunneling, 
\begin{equation}
|S\rangle  = (1-\beta^2)|s\rangle + \sqrt{2}\beta|ex_r\rangle,~
|T0\rangle  =  |t_0\rangle,~ |T\pm\rangle=|t_{\pm}\rangle .
\label{1.3}
\end{equation}
As a result $\Gamma_S/\Gamma_T=1-2\beta^2 <1.$
Due to this relation
the scaling trajectories $E_\Lambda(D)$  determined by the scaling
invariants for eq. (\ref{3.6}) may intersect at a point $D_c$ estimated as
\begin{equation}
\frac{\Gamma_T-\Gamma_S}{\pi}\ln\frac{\pi D_0}{D_c}
=E_T(D_0)-E_S(D_0) \equiv \delta_0. 
\label{3.6a}
\end{equation}
According to numerical calculations of Ref. \cite{KA01}, this level crossing
can occur either before or after the crossover to a Schrieffer-Wolff
regime when the one-electron energies $E_\Lambda(D) - E_{1b}$ exceed
the half-width of reduced continuum band,
$\bar{D}\sim |E_\Lambda(\bar{D}) - E_{1b}|$. 
In both cases the charge degrees of freedom are quenched for
the excitation energy within the interval $-\bar{D}<\varepsilon <\bar{D}$. 
Then, integrating out the residual charge excitations (tunneling transitions
to the states with one and three electrons in the DQD) one comes to an 
effective Hamiltonian including only spin degrees of freedom,
\begin{equation}
\widetilde{H}  =  \sum_{\Lambda=S,T} \bar{E}_\Lambda X^{\Lambda\Lambda}
+\sum_{\langle k\rangle \sigma}\epsilon_k c^\dagger_{k\sigma}c_{k\sigma}
 +  \sum_{\Lambda\Lambda'\lambda}\sum_{kk'\sigma\sigma'}
J^{\Lambda\Lambda'}
X^{\Lambda\Lambda'}c^\dagger_{k\sigma}c_{k'\sigma'}
\label{3.9}
\end{equation}
where $\bar{E}_\Lambda =E_\Lambda(\bar{D})$,
$
J^{\Lambda\Lambda'}\approx 2\sqrt{|\Gamma_\Lambda \Gamma_{\Lambda'}|}
/(\varepsilon_F-\bar{\epsilon}_l)$, 
$X^{\Lambda\Lambda'}=|\Lambda\rangle\langle\Lambda'|$ is a configuration change
operator, $\bar{\epsilon}_l=\epsilon_l(\bar{D})$ 
includes the above RG operation, $\langle k\rangle$ are the states inside
the reduced conduction band.

Unlike conventional Schrieffer-Wolff Hamiltonian for a local spin that 
arises as a result of RG procedure \cite{Hew93}, 
the effective Hamiltonian (\ref{3.9})
contains also the singlet term with renormalized energy 
$\bar{E}_S$ and the effective interaction with coupling constant 
$J^{ST}$ intermixing singlet and triplet states of DQD. The coupling 
constants are connected by the relations $J^S=\alpha^2 J^T,\;\; 
J^{ST}=\alpha J^T~ (\alpha=1-\beta^2)$. This means that the Heitler-London
type DQD possesses the symmetry of a {\it spin rotator}, i.e. that its algebra
$o_4$ is described by two vectors ${\bf S}$ and ${\bf P}$ with 
spherical components
\begin{eqnarray}
S^+  = \sqrt{2}\left(X^{10}+X^{0,-1}\right),~ 
S^-  =  \sqrt{2}\left(X^{01}+X^{-1,0}\right),~
S_z  =  X^{11}-X^{-1,-1},\nonumber\\
P^+  = \sqrt{2}\left(X^{1S}-X^{S,-1}\right),~
P^-  = \sqrt{2}\left(X^{S1}-X^{-1,S}\right),~
P_z  =  -\left(X^{0S}+X^{S0}\right)
\label{3.9c}
\end{eqnarray}
obeying the commutation relations
\[
[S_j,S_k]  = ie_{jkl}S_l,~[P_j,P_k]=ie_{jkl}S_l,~ [P_j,S_k]=ie_{jkl}P_l
\]
($j,k,l$ are Cartesian coordinates). 
These vectors are orthogonal, ${\bf S\cdot P} = 0,$ and the Casimir operator
is ${\bf S}^2+ {\bf P}^2 =3.$
In terms of these operators the interaction term in the effective Hamiltonian 
acquires a symmetric form,
\begin{equation}
\widetilde{H}_{int}=J^T ({\bf S\cdot s})+ J^{ST}\left({\bf P}\cdot{\bf s}\right) 
+\frac{J_T}{2}\sum_{\mu\sigma}X^{\mu\mu}n_{\sigma}, 
\label{3.14}
\end{equation}
where the local electron operators are defined as
\[
n_\sigma=c^\dagger_\sigma c_\sigma=\sum_{kk'} c^\dagger_{k\sigma} c_{k\sigma},\;\;  
{\bf s}=2^{-1/2}\sum_{kk'}\sum_{\sigma\sigma'}
c^\dagger_{k\sigma}\hat{\tau}c_{k'\sigma'}.
\]
($\hat{\tau}$ is the Pauli matrix).
\section{Kondo tunneling}
If the inequality $\bar{D}< D_c$ is satisfied, the triplet becomes the lowest
state of the DQD, i.e., $\bar{\delta}=\bar{E}_T- \bar{E}_S <0$, 
and a Kondo screening for $S=1$ should emerge. However, the
Anderson-type scaling equations for the parameters $J(D)$ involve both 
$J^T$ and $J^{ST}$. In dimensionless variables 
$j_1=\rho_0J^T, j_2=\rho_0J^{ST}, d=\rho_0D$ these equations are written as
\begin{equation}
dj_1/d\ln d = -\left[(j_1)^2+(j_2)^2\right],~~
dj_2/d\ln d = -2j_1j_2. 
\label{3.13}
\end{equation}

Such quasi degeneracy of triplet and singlet states was discussed previously
in relation to the physics of tunneling through 
vertical quantum dots \cite{Gita00}, planar QD occupied by
even number of electrons \cite{Pust00} and strongly asymmetric DQD containing
two dots of essentially different radii \cite{KA01}. In the first two cases 
the singlet-triplet degeneracy is induced by an external magnetic field,
while in the third case the asymmetry is an intrinsic property of the DQD. 
In the case under consideration here, 
the desired asymmetry is induced by the asymmetric 
gate voltage. Analysis of the scaling equations (\ref{3.13}) shows that 
both ${\bf S}$ and ${\bf P}$ operators are involved in anomalous Kondo 
scattering in a case when $\bar{\delta}<T_K$. Then the scaling equations
(\ref{3.13}) are reduced to a single equation for the 
reduced exchange parameter $j_+=j_1+j_2$
\begin{equation}
dj_+/d\ln D=-(j_+)^2
\label{3.18}
\end{equation}
with a fixed point at $j_{+}=\infty$ and the Kondo temperature 
$T_{K0}=\bar{D}\exp(-1/j_+)$. This is a Kondo temperature of a spin rotator,
having the rotational symmetry $SO(4)$. The spin symmetry of DQD is reduced 
together with the Kondo temperature
in the case  $\bar{\delta}\gg T_{K}$. Then the singlet-triplet coupling
$j_2$ is quenched at energies $D\sim\bar{\delta}$, so the low-energy
spectrum is determined by the $S=1$ triplet state. 
The Kondo temperature at large $\bar{\delta}$ is a function
of $\bar{\delta}$ which obeys the law 
$T_K/T_{K0}=(T_{K0}/\bar{\delta})^\lambda,$ where $\lambda$ 
is a universal numerical constant \cite{Pust00}. 

The Kondo-type zero bias anomaly in conductance
arises when $\bar{\delta}<0$ and the DQD is in a state with S=1. It 
 growth with temperature
as $G \sim \ln^{-2}[T/T_K(\bar{\delta})]$ for 
$|\bar{\delta}|\gg T \gg T_K(\bar{\delta})$ and as $G \sim \ln^{-2}[T/T_{K0}]$
for $T\gg T_{K0}\gg \bar{\delta}$. At $T\to 0$ the conductance tends to
the unitarity limit $G_0=2e^2/\pi\hbar$. 

Thus one arrives at the following picture of rearrangement of the low-energy
spectrum of DQD occupied by two electrons in a parallel geometry 
(Fig.1) under external bias. In a symmetric case $V_g^l=V_g^r$ the isolated 
system possesses an axial symmetry and a singlet ground state (Heitler-London 
type artificial molecule). Its spin spectrum consists of 
singlet-triplet excitations, and the singlet state is stabilized by the
indirect exchange energy $E_T-E_S=V^2/Q$ (Fig. 3a). 
Two singlet charge excitons 
(even and odd) corresponding to symmetric and asymmetric combination of 
the polar states $d^\dagger_{i\uparrow}d^\dagger_{i\downarrow}|0\rangle$ 
$(i=l,r)$
are separated by a large energy gap $Q$ from the spin exciton.  The 
level renormalization (\ref{3.6}) is the same for the singlet and triplet
states because the charge transfer excitations are axially symmetric in
this case. The only possibility to open a Kondo channel is 
achievable by a trivial 
switching off the interdot tunneling $(V\to 0)$. Then the singlet-triplet
gap also tends to zero, and when it becomes less than $T_K$ for the left dot,
the latter behaves as conventional $S=1/2$ QD with odd occupation. 
The axial symmetry is broken when $V_g^l-V_g^r>0$, and the "right" 
exciton $E_R$ then softens (Fig. 3b). 
\begin{figure}[htb]
\centering
\includegraphics[
height=0.4\textheight,
keepaspectratio]{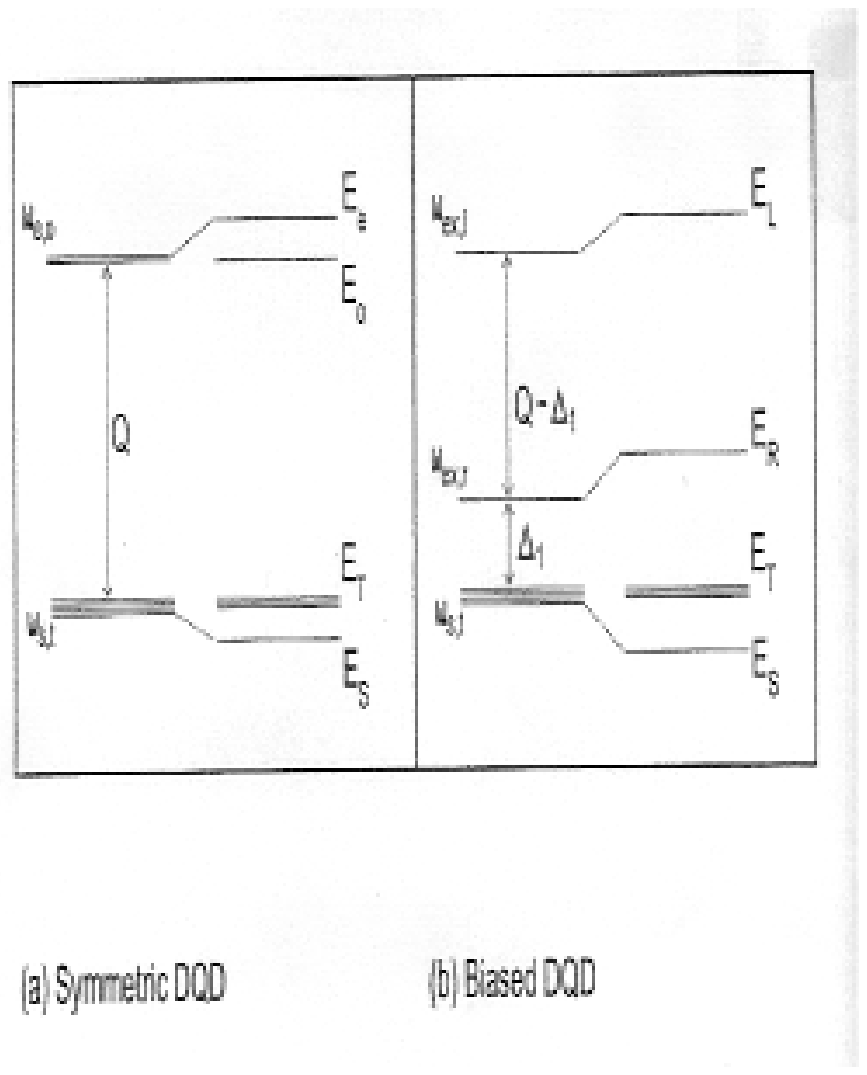}
\caption{Electron levels of double quantum dot at zero ($a$) and 
non-zero ($b$) value of $V_{g}^{r}-V_{g}^{l}$.}
\end{figure}
The system acquires the symmetry of a spin rotator
described by the semi-simple Lie group $SO(4)$ with triplet exciton mixed
with the singlet ground state due to the tunneling interaction with the leads.    
When the condition (\ref{3.6a}) and the inequality
$\bar{D} < D_c$ are valid, the ground state of DQD becomes a triplet
and it behaves as an under-screened $S=1$ Kondo center with zero bias anomaly
of the conductance. Thus, the charge excitations in the right dot that is  
not involved directly in the source-drain tunneling  can drive the spin
excitations in the left dot and induce the Kondo resonance in a DQD with
even occupation. 
%%%%%%%%%%%%%%


\begin{thebibliography}{99}
%
\bibitem{Gogo99} D. Goldhaber-Gordon {\it et.al.,} Phys. Rev. Lett. {\bf81}, 
5225 (1998); S.M. Cronenwett {\it et al.,}, Science, {\bf281}, 540 (1998);
F. Simmel {\it et al.,} Phys. Rev. Lett. {\bf 83}, 804 (1999).
\bibitem{Glazr88b}  L.I. Glazman and M.E. Raikh, JETP Lett. {\bf 47},
452, (1988); T.K. Ng and P.A. Lee, Phys. Rev. Lett., {\bf 61}, 1768 (1988).
\bibitem{Rok00a} L.P. Rokhinson, et al., Phys. Rev. B {\bf60}, 16319 (2000);
J. Schmid, et al., Phys. Rev. Lett. {\bf84}, 5824 (2000).  
\bibitem{APK} M. Pustilnik, Y. Avishai and K. Kikoin, Phys. Rev. Lett. {\bf84}, 
1756 (2000). 
\bibitem{Pust00} M. Pustilnik and L. Glazman, Phys. Rev. Lett. {\bf85}, 2993 (2000);
cond-mat/0102458.
\bibitem{Cobd00} J. Nyg\aa rd, D.H. Cobden and P.E. Lindelof, 
Nature {\bf408}, 342 (2000). 
\bibitem{KA01} K. Kikoin and Y. Avishai, Phys. Rev. Lett. {\bf 86}, 2090 (2001).
\bibitem{Liu98} W. Liu et al, J. Chem. Phys. {\bf107}, 3584 (1997).
\bibitem{Molen95} L.W. Molenkamp et al, Phys. Rev. Lett. {\bf75}, 4282 (1995); 
F. Hofmann et al, Phys. Rev. B{\bf51}, 13872 (1995).
\bibitem{Hew93} A.C. Hewson, {\it The Kondo Problem to Heavy Fermions}
(Cambridge University Press, Cambridge) 1993
\bibitem{Gita00} D. Giuliano and A. Tagliacozzo, Phys. Rev. Lett. {\bf 84}, 
4677 (2000); M. Eto and Yu.V. Nazarov, Phys. Rev. Lett. {\bf85}, 1306 (2000).
\end{thebibliography}
\end{document}